# Uniaxial negative thermal expansion in an orthorhombic superconductor CoZr$_3$

Hiroto Arima[1][‡], Tatsuki Inui[1][‡], Aichi Yamashita[1], Akira Miura[2], Hiroaki Itou[2], Chikako Moriyoshi[3], Hiroshi Fujihisa[4], Yoshikazu Mizuguchi[1]*

[1] Department of Physics, Tokyo Metropolitan University, Hachioji, Tokyo 192-0397, Japan
[2] Faculty of Engineering, Hokkaido University, Sapporo, Hokkaido, 060-8628, Japan
[3] Graduate School of Advanced Science and Engineering, Hiroshima University, Higashihiroshima, Hiroshima, 739-8526, Japan
[4] National Metrology Institute of Japan (NMIJ), National Institute of Advanced Industrial Science and Technology (AIST), Tsukuba, Ibaraki 305-8565, Japan

*Corresponding author: mizugu@tmu.ac.jp
‡ Authors contributed equally to this work

We investigated the temperature evolution of crystal structure of orthorhombic CoZr$_3$, which is a superconductor with a transition temperature of 4.3 K, by synchrotron and laboratory (CuKα) X-ray diffraction. Uniaxial negative thermal expansion along the *c*-axis, which is similar to that observed in tetragonal CoZr$_2$, has been observed at a wide temperature range of *T* = 90‑800 K in CoZr$_3$, while *a*-and *b*-axis exhibit positive thermal expansion.

Negative thermal expansion (NTE) [1–5] has been studied in pure science fields because the emergence of NTE is correlated with various electronic or magnetic ordering, phase transition, or structural instability [5–13]. In addition, NTE and zero thermal expansion (ZTE) materials have been extensively studied for various applications, including precision instruments. A typical NTE material is ZrW$_2$O$_8$, which exhibits a large isotropic NTE over a wide temperature (*T*) range [4]. Furthermore, uniaxial NTE has been observed in some materials, where one axis direction only exhibits NTE and the others exhibit positive thermal expansion (PTE) [14,15]. Recently, we reported uniaxial NTE (and ZTE) in a tetragonal (CuAl$_2$-type) superconductor CoZr$_2$ and related *Tr*Zr$_2$ systems (*Tr*: transition metal) [see Fig. 1(b)] [16–19]. The possible causes of uniaxial NTE in *Tr*Zr$_2$ are large *c*/*a* ratio [17] and flexible Co-Zr bond angle [16]. Volume ZTE can be achieved in a material that shows uniaxial NTE and PTE by compensation of the axis thermal expansions. In $Tr$Zr$_2$, such phenomena would occur in superconducting (metallic) materials. Therefore, further developments of $Tr$-Zr materials that exhibit uniaxial NTE are desired. Here, we report the observation of uniaxial NTE in orthorhombic (Re$_3$B-type) CoZr$_3$ [Fig. 1(a)] with a $Cmcm$ (#63) space group. CoZr$_3$ is a superconductor with a transition temperature ($T_c$) of 4.3 K, which is consistent with a previous report [20].

Polycrystalline samples of CoZr$_3$ were synthesized by arc melting. Co (99%, Kojundo Kagaku) powders and Zr (99.2%, Nilaco) plates with a nominal composition of CoZr$_3$ were melted on a water-cooled Cu stage in an Ar filled arc furnace, and the melting was repeated three times for homogenization. Sample #1 contained small impurity phase (0.6% mass fraction to the main phase) of unreacted Zr ($P6_3/mmc$, #194), and sample #2 was single-phase. Both samples showed superconductivity and uniaxial NTE. Crystal structure of sample #1 was investigated by synchrotron X-ray diffraction (SXRD) at BL02B2, SPring-8 (proposal no.: 2022B1054). The SXRD experiment was performed at temperatures ($T$) of 90–800 K where the sample was sealed in an evacuated quartz capillary, and the sample temperature was controlled by N$_2$ gas. The SXRD data were collected using multiple MYTHEN system with X-ray of $\lambda = 0.49629$ Å [21] with a $2\theta$ step of 0.006 deg. Laboratory XRD with CuK$\alpha$ radiation were performed for sample #2 on Miniflex600 (RIGAKU) equipped with a temperature controller BTS500. Lattice parameters were determined by Rietveld refinement using RIETAN-FP [22], and the crystal structure images were drawn using VESTA [23]. The temperature (and field) dependences of magnetic susceptibility ($4\pi\chi$) (magnetization, $M$) were measured using a superconducting quantum interference devise (SQUID) (MPMS3, Quantum Design) after both zero-field cooling (ZFC) and field cooling (FC).

Figs. 1(a) and 1(b) display the crystal structure images of CoZr$_3$ and CoZr$_2$ depicted along different axis directions. In spite of different structural types, orthorhombic-Re$_3$B and tetragonal-CuAl$_2$ type, both compounds exhibit uniaxial NTE along the $c$-axis. Fig. 1(c) shows the SXRD pattern for sample #1 (upper panel) zoomed at $2\theta = 11.5$–13.5 deg. and the intensity-$T$ map of the region at $T = 100$–800 K (lower panel). As indicated by a square, the 004 peak exhibits shift to a lower angle, which indicates expansion of the $c$-axis, with decreasing temperature.

To clarify the temperature evolutions of lattice parameters, the $T$-scan SXRD patterns and laboratory XRD patterns (Miniflex600) were analyzed by the Rietveld method. An example of the refinement of SXRD pattern is shown in Fig. 2(e), where two-phase analysis was performed with an impurity phase of Zr ($P6_3/mmc$, 0.6% mass fraction to CoZr$_3$). Similar reliability factors were obtained for analyses of the patterns taken at 90–600 K. On the SXRD patterns at $T > 600$ K, we found that intensity of some peaks is enhanced due to possible grain connectivity [Fig. S1 (supplemental materials [24])], but the orthorhombic *Cmcm* structure maintains up to the highest temperature reported in this paper. We confirmed that the *Cmcm* structure of CoZr$_3$ is more stable than a symmetry-lowered *Cmc*$2_1$ structure and a lattice-distorted monoclinic *P*$2_1$ structure by DFT calculations (CASTEP code[25]). In Fig. S2 (supplemental materials [24]), we show the Rietveld refinement result for laboratory XRD pattern ($T = 552$ K). The lattice parameters determined from laboratory XRD are plotted in Figs. 2(a–d) together with those obtained from SXRD analyses. Although *a*, *b*, and *V* exhibit PTE, a clear uniaxial NTE is observed for *c*. From two diffractometers with different wave lengths (on different-batch samples), we observed uniaxial NTE along the *c*-axis for CoZr$_3$.

Superconducting properties of CoZr$_3$ have been examined by magnetic measurements. Fig. 3 shows the temperature dependence of magnetic susceptibility ($4\pi\chi$). A large shielding signal was observed for ZFC data. The difference between ZFC and FC is typical behavior of a type-II superconductor. $T_c$ estimated from $4\pi\chi$ is 4.3 K, which is slightly higher than $T_c$ reported by a previous report ($T_c = 3.55$ K) [20], but the emergence of bulk superconductivity is consistent. In Figs. S3 and S4 (supplemental materials [24]), *M-T* data under various magnetic fields and estimated upper critical field ($B_{c2}$)-*T* phase diagram is shown. The estimated $B_{c2}$ ($T = 0$ K) is 2.9 T. Above $T_c$, no magnetic transition was observed at $T < 300$ K. The emergence of bulk superconductivity and the absence of magnetic ordering at low temperatures were confirmed by *M-H* loops [Fig. S6 (supplemental materials [24])].

In conclusion, we synthesized polycrystalline samples of orthorhombic (Re$_3$B-type) CoZr$_3$ by arc melting and investigated the structural and magnetic properties. From temperature-dependent XRD (SXRD and laboratory XRD), uniaxial NTE along the *c*-axis has been confirmed for CoZr$_3$. Since the emergence of uniaxial NTE is common for CoZr$_2$ and CoZr$_3$, low-dimensional crystal structure (tetragonal, orthorhombic, etc.) containing Co-Zr bond would be essential for the uniaxial NTE in those systems. The next step is to identify the cause of uniaxial NTE by analyzing phonon dispersions on those materials, which will allow us to further design new NTE materials. In addition, as suggested in *Tr*Zr$_2$ systems [16], tuning of PTE along the *a*- or *b*-axis and NTE along the *c*-axis will achieve volume ZTE in CoZr$_3$-based materials.

**Acknowledgment**

The authors would like to thank T. Yagi and O. Miura for their support with the experiments and discussions. This work was partly supported by a Grant-in-Aid for Scientific Research (KAKENHI) (21K18834, 21H00151), JST-ERATO (JPMJER2201), and the Tokyo Government Advanced Research (H31-1).

*mizugu@tmu.ac.jp

1) K. Takenaka, Sci. Technol. Adv. Mater. 13, 013001 (2012).
2) G. D. Barrera, J. A. O. Bruno, T. H. K. Barron, and N. L. Allan, J. Phys.: Condens. Matter 17, R217 (2005).
3) J. Chen, L. Hu, J. Deng, and X. Xing, Chem. Soc. Rev. 44, 3522 (2015).
4) T. A. Mary, J. S. O. Evans, T. Vogt, and A. W. Sleight, Science 272, 90 (1996).
5) K. Takenaka, Y. Okamoto, T. Shinoda, N. Katayama, and Y. Sakai, Nat. Commun. 8, 14102 (2016).
6) M. Braden, G. André, S. Nakatsuji, and Y. Maeno, Phys. Rev. B58, 847 (1998).
7) R. Huang, Y. Liu, W. Fan, J. Tan, F. Xiao, L. Qian, and L. Li, J. Am. Chem. Soc. 135, 11469 (2013).
8) I. Yamada, K. Tsuchida, K. Ohgushi, N. Hayashi, J. Kim, N. Tsuji, R. Takahashi, M. Matsushita, N. Nishiyama, T. Inoue, T. Irifune, K. Kato, M. Takata, and M. Takano, Angew. Chem. Int. Ed. 50, 6579 (2011).
9) M. Azuma, W. Chen, H. Seki, M. Czapski, S. Olga, K. Oka, M. Mizumaki, T. Watanuki, N. Ishimatsu, N. Kawamura, S. Ishiwata, M. G. Tucker, Y. Shimakawa, J. P. Attfield, Nat. Commun. 2, 347 (2011).
10) J. R. Salvador, F. Guo, T. Hogan, M. G. Kanatzidis, Nature 425, 702 (2003).
11) T. Yokoyama and K. Eguchi, Phys. Rev. Lett. 107, 065901 (2011).
12) J. Chen, X. R. Xing, G. R. Liu, J. H. Li, and Y. T. Liu, Appl. Phys. Lett. 89, 101914 (2006).
13) X. R. Xing, J. X. Deng, J. Chen, and G. R. Liu, Rare Metals 22, 294 (2003).
14) F. H. Gillery and E. A. Bush, J. Am. Ceram. Soc. 42, 175 (1959).
15) C. Ablitt, S. Craddock, M. S. Senn, A. A. Mostofi, and N. C. Bristowe, npj Comp. Mater. 3, 44 (2017).

16) Y. Mizuguchi, Md. R. Kasem, and Y. Ikeda, J. Phys. Soc. Jpn. 91, 103601 (2022).

17) H. Arima, Md. R. Kasem, and Y. Mizuguchi, arXiv: 2210.10367.

18) Md. R. Kasem, H. Arima, Y. Ikeda, A. Yamashita, and Y. Mizuguchi, J. Phys.: Mater. 5, 045001 (2022).

19) Y. Watanabe, H. Arima, H. Usui, and Y. Mizuguchi, 2211.00958.

20) T. Koyama, S. Ogura, S. Ikeda, T. Momura, K. Ueda, T. Mito, and T. Kohara, JPS meeting abstract (September meeting 2013), 26aPS −145, available at https://www.jstage.jst.go.jp/article/jpsgaiyo/68.2.3/0/68.2.3_535_3/_pdf/-char/ja.

21) S. Kawaguchi, M. Takemoto, K. Osaka, E. Nishibori, C. Moriyoshi, Y. Kubota, Y. Kuroiwa, and K. Sugimoto, Rev. Sci. Instrum. 88, 085111 (2017).

22) F. Izumi and K. Momma, Solid State Phenom. 130, 15 (2007).

23) K. Momma and F. izumi, J. Appl. Crystallogr. 41, 653 (2008).

24) Supplemental materials: Rietveld refinement results and magnetization data are shown.

25) S. J. Clark, M. D. Segall, C. J. Pickard, P. J. Hasnip, M. I. J. Probert, K. Refson, and M. C. Payne, Zeitschrift für Kristallographie - Crystalline Materials 220, 567 (2005).

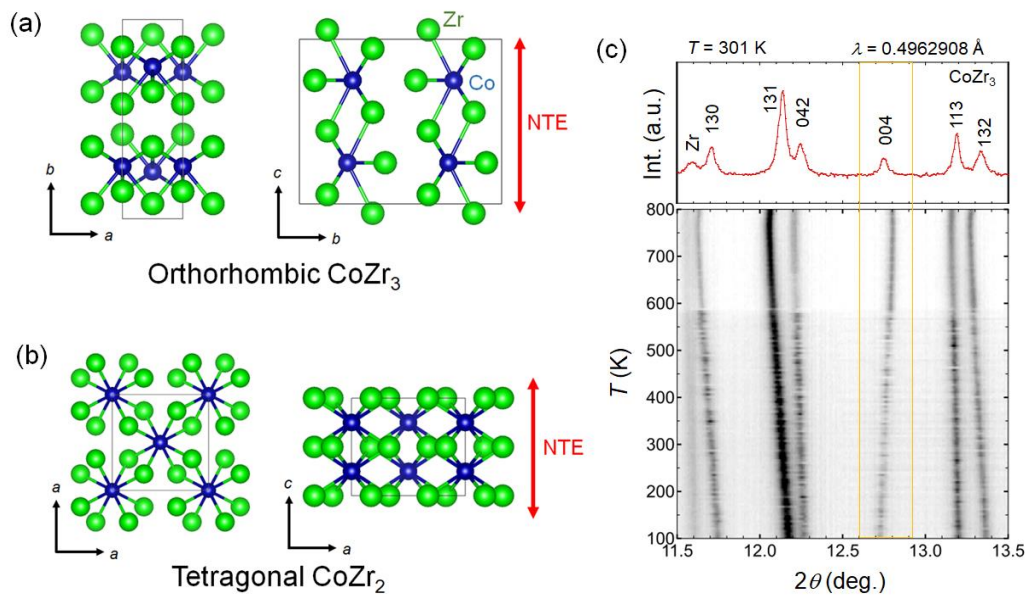

Fig. 1. (Colour online) (a,b) Schematic images of crystal structure of orthorhombic (Re$_3$B-type) CoZr$_3$ and tetragonal (CuAl$_2$-type) CoZr$_2$. Square lines indicate their unit cell. (c) The upper panel is a SXRD pattern ($T$ = 301 K) zoomed at $2\theta$ = 11.5–13.5 deg. and the lower panel is the intensity (Int.)-$T$ map of the region at $T$ = 100–800 K.

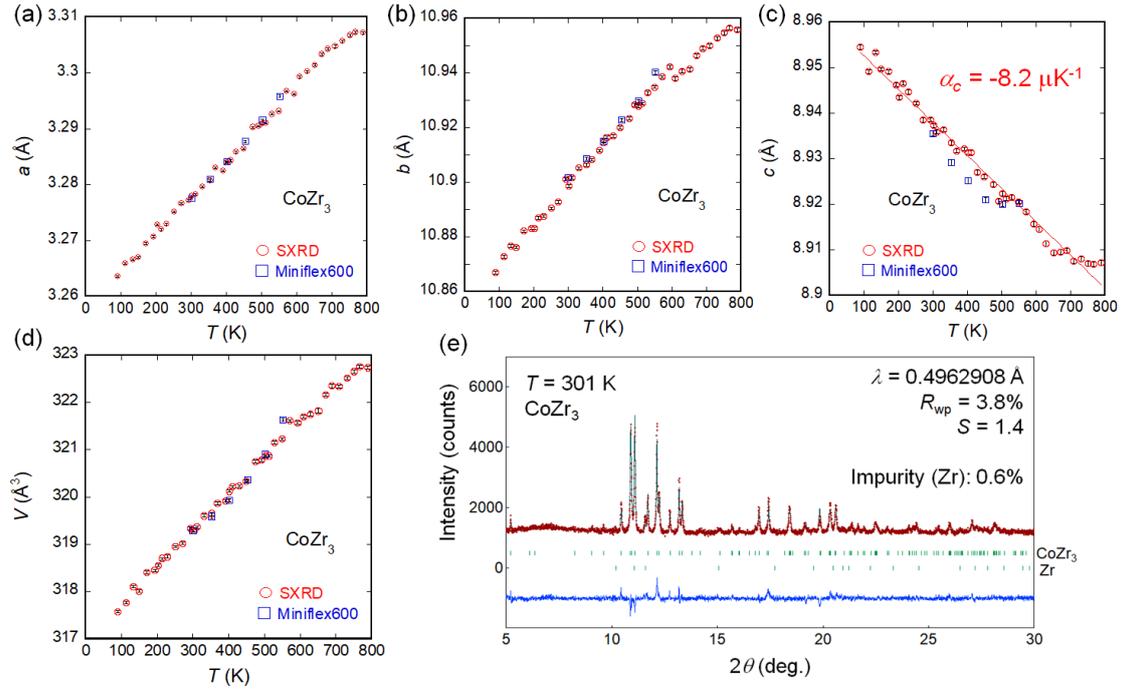

Fig. 2. (Colour online) (a–d) Temperature dependences of lattice parameters of $a$, $b$, $c$, and $V$. The SXRD data were taken on sample #1, and the Miniflex600 data were taken on sample #2. The linear line in (c) indicates linear fitting results for the SXRD data, and the thermal expansion constant was estimated as $\alpha_c = -8.2\ \mu K^{-1}$. (e) Rietveld refinement result for SXRD data ($T = 301$ K).

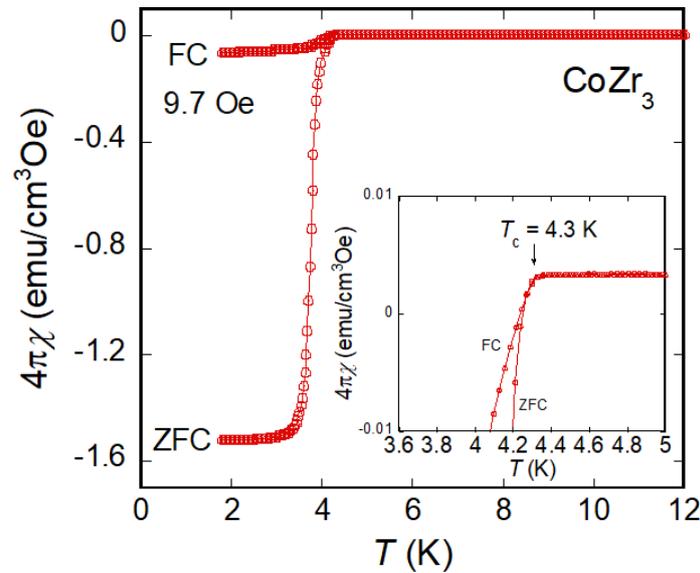

Fig. 3. (Colour online) Temperature dependence of magnetic susceptibility for $CoZr_3$ (sample #2). The inset shows data around $T_c$.

# Supplemental Materials

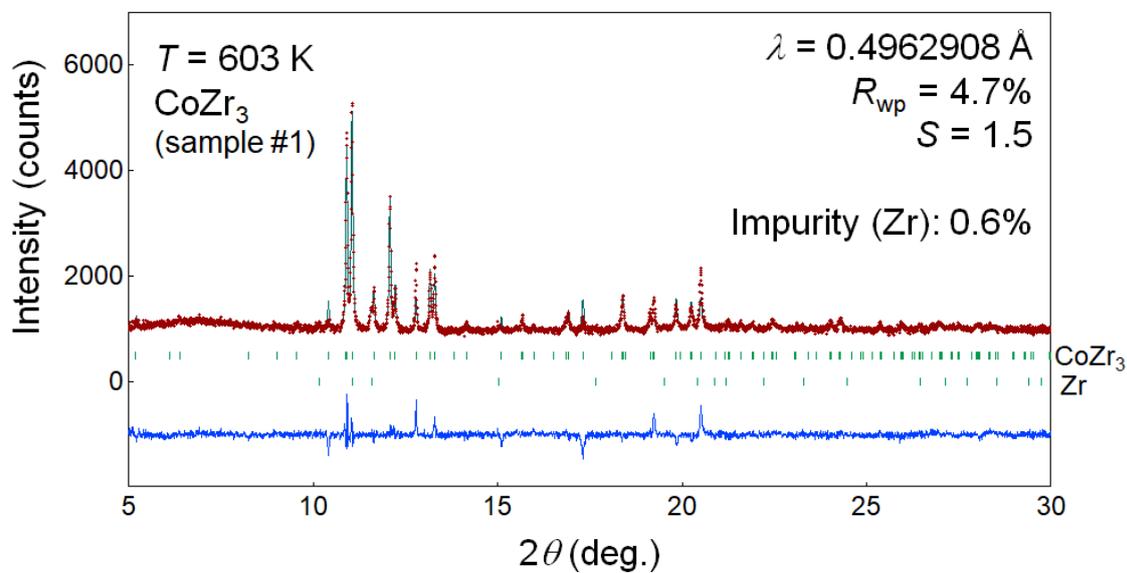

Fig. S1. Rietveld refinement result for SXRD pattern of CoZr$_3$ (sample #1, $T$ = 603 K).

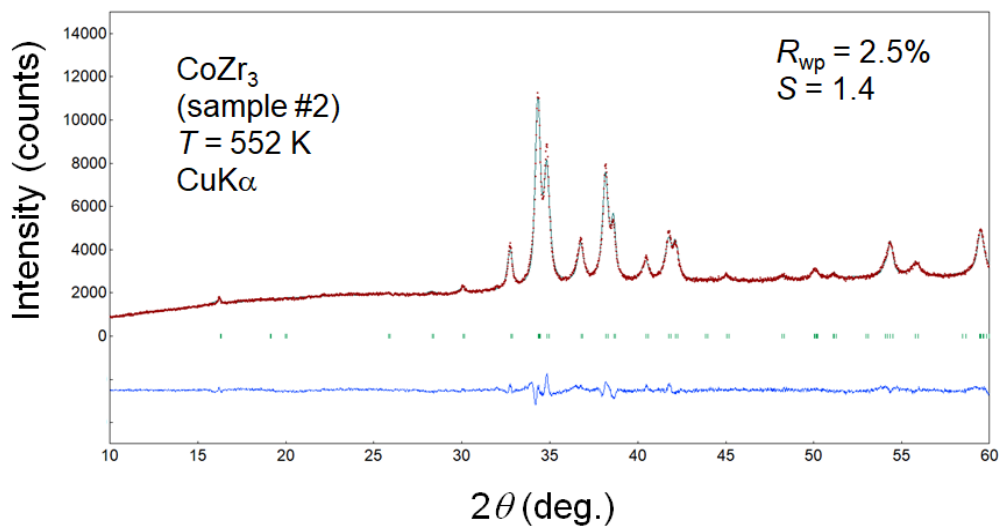

Fig. S2. Rietveld refinement result for laboratory XRD pattern of CoZr$_3$ (sample #2, $T$ = 552 K).

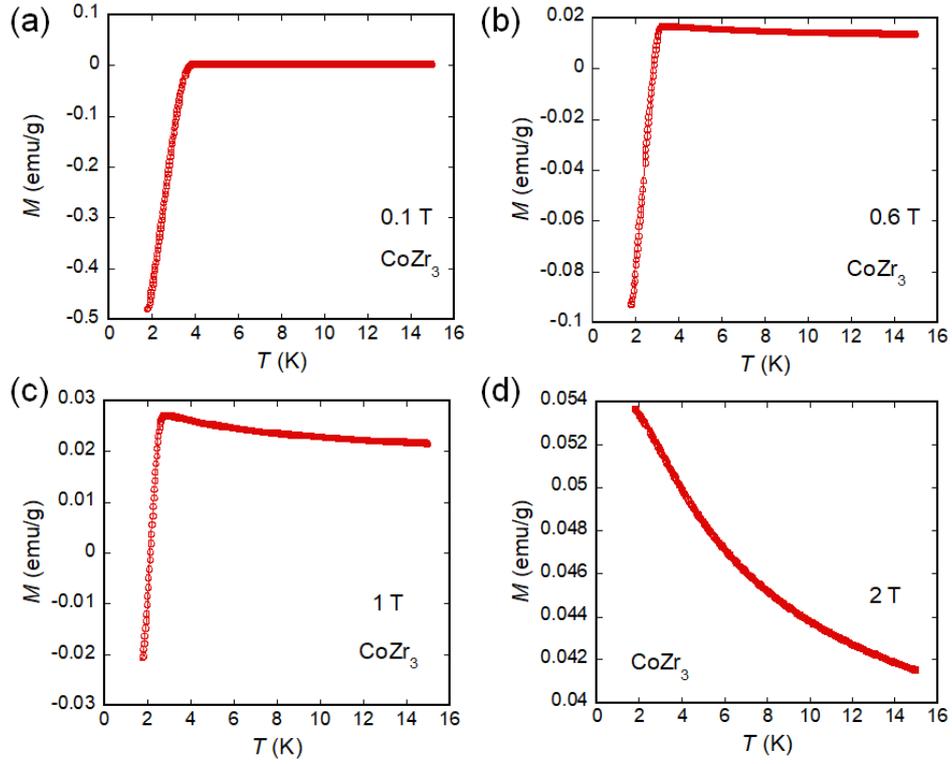

Fig. S3. Temperature dependences of magnetization ($M$) for CoZr$_3$ (sample #2) under magnetic fields. The magnetization measurements were performed after zero-field cooling.

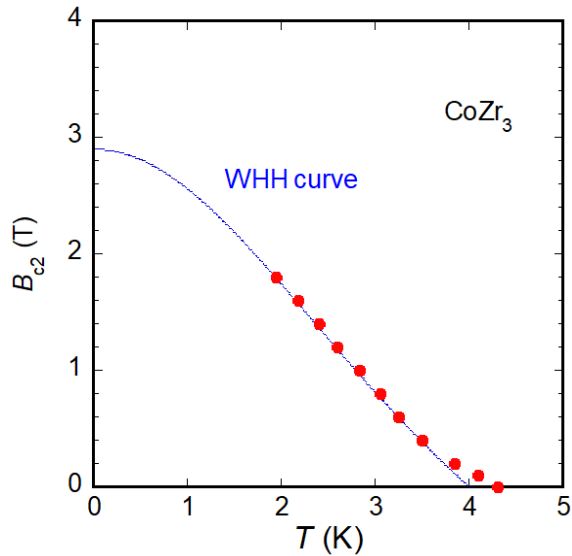

Fig. S4. Temperature dependence of upper critical field ($B_{c2}$) estimated from the field dependence of $T_c$ (magnetization). The curve shows the WHH model [S1] explaining the data at $T < 3.5$ K.

[S1] N. R. Werthamer, E. Helfand, and P. C. Hohenberg, Phys. Rev. 147, 295 (1966).

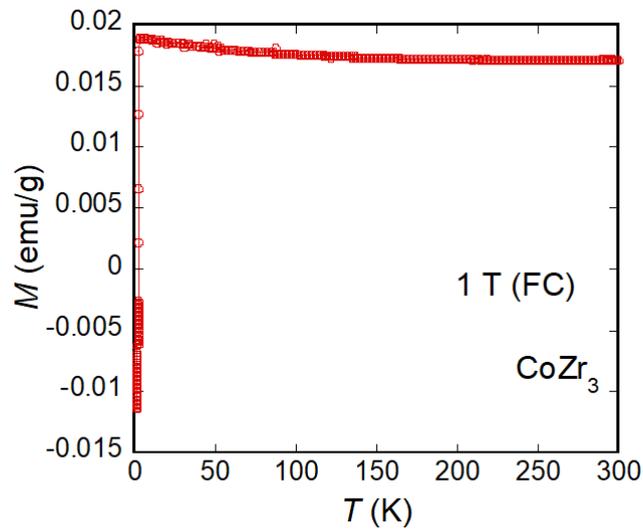

Fig. S5. Temperature dependences of $M$ for CoZr$_3$ (sample #1) under 1 T (field-cooled data). No magnetic transition was observed at $T_c < T < 300$ K.

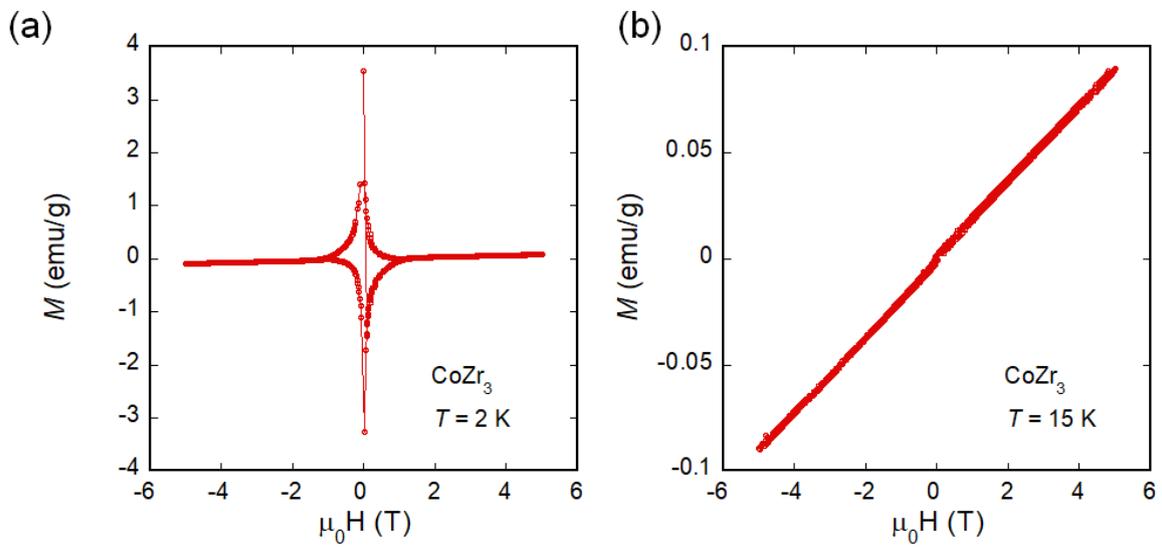

Fig. S6. M-H curves for CoZr$_3$ (sample #1) at temperatures below $T_c$ ($T = 2$ K) and above $T_c$ ($T = 15$ K)